\documentstyle[12pt]{article}
\begin{document}
\hbadness=10000
\hbadness=10000
\begin{titlepage}
\nopagebreak
\begin{flushright}
{\normalsize
HIP-1998-25/TH\\
BONN-TH-98-10\\
KANAZAWA-98-05\\
May, 1998}\\
\end{flushright}
\vspace{0.5cm}
\begin{center}

{\large \bf  Constraints on Supersymmetric $SO(10)$ GUTs \\
with Sum Rules among Soft Masses}

\vspace{0.8cm}

{\bf Yoshiharu Kawamura $^{a,b}$ 
\footnote[1]{e-mail: kawamura@dirac.physik.uni-bonn.de, Humboldt Fellow}}, 
{\bf Tatsuo Kobayashi $^{c,d}$
\footnote[2]{e-mail: kobayash@rock.helsinki.fi}}\\
and\\
{\bf Hitoshi Shimabukuro $^e$
\footnote[3]{e-mail: simabuku@hep.s.kanazawa-u.ac.jp}}

\vspace{0.5cm}
$^a$ Physikalisches Institut, Universit\"at Bonn \\
   D-53115 Bonn, Germany, \\
$^b$ Department of Physics, Shinshu University \\
   Matsumoto 390, Japan, \\
$^c$ Department of Physics, High Energy Physics Division\\
    University of Helsinki, \\
$^d$ Helsinki Institute of Physics\\
    P.O. Box 9, FIN-00014 Helsinki, Finland \\
and\\
$^e$ Department of Physics, Kanazawa University\\
Kanazawa, 920-1192, Japan\\

\end{center}
\vspace{0.5cm}

\nopagebreak

\begin{abstract}
We study phenomenological aspects of supersymmetric $SO(10)$ GUTs 
with sum rules among soft SUSY breaking parameters.
In particular, the sum rule related to the stau mass leads to 
the constraints from the requirements of 
successful electroweak breaking and 
the positivity of stau mass squared.
The bottom quark mass is also estimated.
\end{abstract}
\vfill
\end{titlepage}
\pagestyle{plain}
\newpage
\def\thefootnote{\fnsymbol{footnote}}

The unification of force based on the minimal supersymmetric standard 
model (MSSM) \cite{SUSY-GUT} has been hopeful from the data of precision 
measurements \cite{LEP}.
The supersymmetric $SO(10)$ grand unified theory (SUSY $SO(10)$ GUT) 
\cite{SO(10)} is one of 
attractive candidates
of realistic theory at and above the GUT scale $M_{X}$ because
a simplest unification of quarks and leptons can be realized in each
family.
After the gauge symmetry breakdown, the remnant exists in the MSSM as 
specific relations among physical parameters at $M_X$, which are
usually used as initial conditions on the analysis by the use of
renormalization group equations (RGEs), e.g., $g_3 = g_2 = g_1 \equiv g$
for the gauge couplings of the SM gauge group $G_{SM} = SU(3)_C \times 
SU(2)_L \times U(1)_Y$ and
$Y_t^{\alpha\beta} = Y_b^{\alpha\beta} = Y_{\tau}^{\alpha\beta} 
\equiv Y^{\alpha\beta}$
for the Yukawa coupling matrices of up-type quarks, down-type quarks
and leptons.
Here $\alpha$, $\beta$ are the family indices.
The number of independent soft SUSY breaking parameters is also
reduced by the $SO(10)$ symmetry, i.e.,
the parameters are given as 
$({m_{16}}^{\alpha}_{\beta}, m_H, M, A^{\alpha\beta}, B)$ at $M_X$ for
soft squark and slepton masses, soft Higgs masses, gaugino masses,
$A$-parameters and $B$-parameter up to the contributions of $SO(10)$
breaking
including $D$-term contribution \cite{Dterm1,Dterm2}.

The magnitude of each parameter is expected to be determined by
some underlying theory or new concept of SUSY-GUT.
The supergravity (SUGRA) is regarded as a promising theory, describes 
the physics beyond SUSY-GUT effectively, and offers an interesting
scenario
of the origin of soft SUSY breaking parameters \cite{SUGRA}.
The structure of the SUGRA model is reflected on the pattern 
of the parameters, 
e.g., the universal soft SUSY breaking parameters originate
in models with a canonical K\"ahler potential.
The analyses based on this type of initial conditions have been
intensively
carried out \cite{univ}. 

There is another interesting scenario to control the relations among 
physical parameters.
The parameters can be reduced by the adoption of a new concept
$\lq$coupling reduction' by the use of RG invariant relations \cite{CR}.
The assumption in a model is that the Yukawa couplings 
are expressed RG-invariantly by the gauge coupling $g$ as
\begin{eqnarray}
&~& Y_{ijk} = g {\rho}_{ijk} , \label{Y-g}
\end{eqnarray} 
where $\rho_{ijk}$ are model-dependent constant independent of $g$
at the tree level.
Higher order corrections are systematically calculated.
It is called gauge-Yukawa unified (GYU) model \cite{GYU}.
Recently, it is found that the following relations of soft parameters 
are RG-invariant\cite{softGYU,softGYU2},\footnote{
We can calculate higher order corrections to RG invariant relations
\cite{all-order}.} 
\begin{eqnarray}
&~& \sum m^2 \equiv m_i^2+m_j^2+m_k^2=M^2 , \label{sumrule1} \\
&~& A_{ijk} = -M \label{sumrule2}
\end{eqnarray}
where $Y_{ijk}$, $m_i$ and $A_{ijk}$ are Yukawa couplings, soft scalar
masses and $A$-parameters.
The indices $i,j,k$ denote the particle species.
It has been known that the relations 
(\ref{sumrule1}) and (\ref{sumrule2}) are derived in other several 
theories, i.e., the finite field theories \cite{JJ}, 
a certain class of 4-dimensional models from superstring theory (SST)
\cite{BIM} and a non-minimal SUGRA model with a certain type of structure
regarded as a generalization of models from SST 
\cite{softGYU}. 
Hence this type of relations can give a hint to high energy physics
beyond the MSSM.
Therefore it is important to study the phenomenological implications
and low-energy consequences from this type of relations 
\cite{softGYU3,B}.

The low-energy constraints from (\ref{sumrule1}) and (\ref{sumrule2})
are studied in Ref. \cite{softGYU2} based on finite SUSY $SU(5)$
GUT models and 
it is shown that eigenvalues of stau masses squared 
$m_{\tilde \tau_{1,2}}^2$ tend to be negative in some parameter region. 
The sbottom and stop fields always have heavier masses than the 
lightest stau.
Thus the condition of the positivity $m_{\tilde \tau_{1,2}}^2 > 0$ 
as well as the electroweak symmetry breaking conditions constrains 
the parameter regions severely in this type of models.

In the sector where the above relations (\ref{Y-g}),
(\ref{sumrule1}) and (\ref{sumrule2}) hold on, independent
parameters are limited to $g$, $M$ and $m_i$.
The radiative corrections of $m_i^2$ are given as functions of
$g$ and $M$ because the contribution from Yukawa couplings contains
soft scalar masses only as a combination of $\sum m^2$.
In this sense, this type of soft SUSY breaking parameters are much 
more restrictive than a general non-universal one.

Let us compare the excluded regions of soft stau mass
$m_{\tilde{\tau}}$
and gaugino mass $M$ at $M_X$
between the models with the universal type of soft masses $(M, m_0)$
inspired by a minimal SUGRA and the models
with the relations (\ref{sumrule1}) and (\ref{sumrule2}) at $M_X$,
based on the MSSM with large $\tan \beta$.
{}From the requirement of successful electroweak
symmetry breakdown, there is the constraint 
$m_{H_1}^2 - m_{H_2}^2 > M_Z^2$ at the weak scale. 
Here $m_{H_1}^2$ and $m_{H_2}^2$ are soft Higgs masses squared 
with the hypercharge $-1/2$ and $1/2$, respectively.
Using the analysis of RGEs, we get the relation such that
$m_{H_1}^2 - m_{H_2}^2 = c_1 M^2 - c_2 \sum m^2/3$ 
under the condition that 
$m_{H_1}^2 = m_{H_2}^2$ at $M_X$ \cite{COPW} 
and the constant factors $c_1$ and $c_2$ are of ${\cal O}(0.2)$. From
these formulae, the region with $m_{\tilde{\tau}} (=m_0) > M$, 
$(m_{\tilde{\tau}} \gg M_Z)$ 
is excluded in the universal case.\footnote{
For the study on case with non-universal initial conditions for
soft SUSY breaking parameters, see Refs. \cite{Nuniver,KK}.}
On the other hand, in the case with sum rules, the excluded region is 
$M < {\cal O}(200)$GeV independent of
the value of $m_{\tilde{\tau}}$. 
Another requirement is the positivity of physical stau mass squared.
The Yukawa coupling induces to a radiative correction with a negative sign
to the mass squared.
Thus the stau mass squared can be negative if the magnitude of 
$\sum m^2$ is sizable in the large $\tan \beta$ scenario.
This happens easily in the case with sum rules because of
the existence of the relation $\sum m^2 = M^2$
and, in this case, the region with $M \gg m_{\tilde{\tau}}$ is
excluded.
In the universal case, the situation is different because 
the contribution including the factor $\sum m^2(=3m_0^2)$ becomes tiny 
when the value of $m_0=m_{\tilde{\tau}}$ is small, i.e. 
$M \gg m_0$.
Hence it is expected that the excluded regions of $(M, m_{\tilde{\tau}})$
are located at the opposite corners (besides $M < {\cal O}(200)$GeV
in the latter case) each other from the above two
phenomenological requirements.

In this paper, we study generic SUSY $SO(10)$ GUT model with sum rules
among soft SUSY breaking parameters and make sure the above estimation
of the parameter regions, quantitatively, 
imposing the conditions of successful
electroweak symmetry breaking and the positivity
of physical stau mass squared. 
The method of analysis is almost same as that made in Ref.
\cite{softGYU2}.
SUSY corrections to the bottom quark mass are also estimated.

First we give a brief review on the derivation 
of the relations (\ref{Y-g}), (\ref{sumrule1}) and (\ref{sumrule2})
in GYU-models.
We assume that parameters $Y_{ijk}$, $m_i^2$ and $A_{ijk}$ are
expressed in terms of $g$ and $M$.
The relations (\ref{Y-g}), (\ref{sumrule1}) and (\ref{sumrule2}) 
are obtained by solving so-called reduction equations
perturbatively,
\begin{eqnarray}
&~&\beta_{Y_{ijk}}=\beta_g {d Y_{ijk} \over dg} ,  \label{R-Y}\\
&~&\beta_{m^2_i}=\beta_M{\partial {m^2_i} \over \partial M}+
\beta_{M^\dagger}{\partial {m^2_i} \over \partial M^{\dagger}}+
\beta_g{\partial {m^2_i} \over \partial g} ,   \label{R-m}\\
&~&\beta_{A_{ijk}}=\beta_M{\partial A_{ijk}\over \partial M}+
\beta_{M^\dagger}{\partial A_{ijk}\over \partial M^{\dagger}}+
\beta_g{\partial A_{ijk}\over \partial g}    \label{R-A}
\end{eqnarray}
where $\beta_X$ denotes a $\beta$-function of parameter $X$.
For the application on an explicit model, see Ref. \cite{GYU2}.

Second we give some basic assumptions and relations on our analysis.
The first assumption is that 
all of quarks and leptons in each family belong to one  
16-plet under $SO(10)$ and
this 16-plet has the Yukawa coupling such as $(16)^2H$ where $H$ 
is 10-plet including $H_1$ and $H_2$.
We ignore the family mixing effects.
The second one is
that the relations (\ref{sumrule1}) and (\ref{sumrule2}) 
hold on at $M_X$ in the third family.
The third one is that there are no extra contributions on the
symmetry breaking $SO(10) \to G_{SM}$.
(Later we relax this assumption by the introduction of 
$D$-term contribution.)
Our initial conditions at $M_X$ are summarized as follows,
\begin{eqnarray}
&~& g_3=g_2=g_1=g ,~~   Y_t=Y_b=Y_\tau=Y ,~ \nonumber\\
&~& m_{\tilde Q}^2=m_{\tilde t}^2=m_{\tilde b}^2
=m_{\tilde \tau_L}^2=m_{\tilde \tau_R}^2=m_{16}^2 , \nonumber\\
&~&m_{H_1}^2=m_{H_2}^2=m_{H}^2  , \nonumber\\
&~& A_t=A_b=A_\tau=-M ,~~   2m_{16}^2+m_H^2=M^2 \label{relations} 
\end{eqnarray} 
where ${\tilde Q}$, ${\tilde t}$, ${\tilde b}$, ${\tilde \tau}_L$ 
and ${\tilde \tau}_R$ denote the $SU(2)_L$
doublet squark of the third family, the singlet stop, 
the singlet sbottom, the stau in the doublet slepton of the third family
and the singlet stau.
Here and hereafter we omit the index representing the third family. 

Next we parametrize the Yukawa coupling $Y$ using $g$ as $Y=\rho g$.
The value of $\rho$ gives an important information on
the matter content and its interactions in GYU-models and/or
the structure of superpotential in SUGRA.
We take the following input parameters, 
\begin{equation}
M_\tau=1.777 {\rm GeV}, \quad M_Z=91.188 {\rm GeV},
\end{equation}
\begin{equation}
\alpha^{-1}_{\rm EM}(M_Z) = 127.9 +{8 \over 9 \pi} 
\log {M_t \over M_Z},
\end{equation}
\begin{equation}
\sin ^2 \theta_W(M_Z) =0.2319 - 3.03\times 10^{-5} T-8.4 \times 
10^{-8}T^2
\end{equation}
where $T=M_t/[{\rm GeV}]-165$.
Here $M_\tau$ and $M_t$ are physical tau lepton and top quark masses.
The Yukawa unification condition gives the predicted top quark mass from
the 
above experimental value of $M_\tau$ for each value of $\rho$.
Fig. 1 shows the predicted value of the physical top quark mass 
for $k \equiv \rho^2$.
Thus we find the realistic region such that $0.7 \leq k \leq 1.4$
to obtain the present experimental value of the top mass,
$M_t=175.6 \pm 5.5 {\rm GeV}$.
For example, the value $k = 1.0$ leads to $M_t=175$ GeV and 
$\tan \beta = 53$, while $\tan \beta = 50$ and 55 for
$k= 0.7$ and 1.4, respectively.
\begin{center}
\setlength{\unitlength}{0.240900pt}
\ifx\plotpoint\undefined\newsavebox{\plotpoint}\fi
\sbox{\plotpoint}{\rule[-0.200pt]{0.400pt}{0.400pt}}%
\begin{picture}(1500,900)(0,0)
\font\gnuplot=cmr10 at 10pt
\gnuplot
\sbox{\plotpoint}{\rule[-0.200pt]{0.400pt}{0.400pt}}%
\put(220.0,113.0){\rule[-0.200pt]{4.818pt}{0.400pt}}
\put(198,113){\makebox(0,0)[r]{160}}
\put(1416.0,113.0){\rule[-0.200pt]{4.818pt}{0.400pt}}
\put(220.0,240.0){\rule[-0.200pt]{4.818pt}{0.400pt}}
\put(198,240){\makebox(0,0)[r]{165}}
\put(1416.0,240.0){\rule[-0.200pt]{4.818pt}{0.400pt}}
\put(220.0,368.0){\rule[-0.200pt]{4.818pt}{0.400pt}}
\put(198,368){\makebox(0,0)[r]{170}}
\put(1416.0,368.0){\rule[-0.200pt]{4.818pt}{0.400pt}}
\put(220.0,495.0){\rule[-0.200pt]{4.818pt}{0.400pt}}
\put(198,495){\makebox(0,0)[r]{175}}
\put(1416.0,495.0){\rule[-0.200pt]{4.818pt}{0.400pt}}
\put(220.0,622.0){\rule[-0.200pt]{4.818pt}{0.400pt}}
\put(198,622){\makebox(0,0)[r]{180}}
\put(1416.0,622.0){\rule[-0.200pt]{4.818pt}{0.400pt}}
\put(220.0,750.0){\rule[-0.200pt]{4.818pt}{0.400pt}}
\put(198,750){\makebox(0,0)[r]{185}}
\put(1416.0,750.0){\rule[-0.200pt]{4.818pt}{0.400pt}}
\put(220.0,877.0){\rule[-0.200pt]{4.818pt}{0.400pt}}
\put(198,877){\makebox(0,0)[r]{190}}
\put(1416.0,877.0){\rule[-0.200pt]{4.818pt}{0.400pt}}
\put(321.0,113.0){\rule[-0.200pt]{0.400pt}{4.818pt}}
\put(321,68){\makebox(0,0){0.4}}
\put(321.0,857.0){\rule[-0.200pt]{0.400pt}{4.818pt}}
\put(524.0,113.0){\rule[-0.200pt]{0.400pt}{4.818pt}}
\put(524,68){\makebox(0,0){0.6}}
\put(524.0,857.0){\rule[-0.200pt]{0.400pt}{4.818pt}}
\put(727.0,113.0){\rule[-0.200pt]{0.400pt}{4.818pt}}
\put(727,68){\makebox(0,0){0.8}}
\put(727.0,857.0){\rule[-0.200pt]{0.400pt}{4.818pt}}
\put(929.0,113.0){\rule[-0.200pt]{0.400pt}{4.818pt}}
\put(929,68){\makebox(0,0){1}}
\put(929.0,857.0){\rule[-0.200pt]{0.400pt}{4.818pt}}
\put(1132.0,113.0){\rule[-0.200pt]{0.400pt}{4.818pt}}
\put(1132,68){\makebox(0,0){1.2}}
\put(1132.0,857.0){\rule[-0.200pt]{0.400pt}{4.818pt}}
\put(1335.0,113.0){\rule[-0.200pt]{0.400pt}{4.818pt}}
\put(1335,68){\makebox(0,0){1.4}}
\put(1335.0,857.0){\rule[-0.200pt]{0.400pt}{4.818pt}}
\put(220.0,113.0){\rule[-0.200pt]{292.934pt}{0.400pt}}
\put(1436.0,113.0){\rule[-0.200pt]{0.400pt}{184.048pt}}
\put(220.0,877.0){\rule[-0.200pt]{292.934pt}{0.400pt}}
\put(45,495){\makebox(0,0){$M_t$(GeV)}}
\put(828,23){\makebox(0,0){$k$}}
\put(220.0,113.0){\rule[-0.200pt]{0.400pt}{184.048pt}}
\multiput(364.59,113.00)(0.488,0.758){13}{\rule{0.117pt}{0.700pt}}
\multiput(363.17,113.00)(8.000,10.547){2}{\rule{0.400pt}{0.350pt}}
\multiput(372.58,125.00)(0.498,0.637){99}{\rule{0.120pt}{0.610pt}}
\multiput(371.17,125.00)(51.000,63.734){2}{\rule{0.400pt}{0.305pt}}
\multiput(423.58,190.00)(0.498,0.560){97}{\rule{0.120pt}{0.548pt}}
\multiput(422.17,190.00)(50.000,54.863){2}{\rule{0.400pt}{0.274pt}}
\multiput(473.00,246.58)(0.520,0.498){95}{\rule{0.516pt}{0.120pt}}
\multiput(473.00,245.17)(49.928,49.000){2}{\rule{0.258pt}{0.400pt}}
\multiput(524.00,295.58)(0.607,0.498){81}{\rule{0.586pt}{0.120pt}}
\multiput(524.00,294.17)(49.784,42.000){2}{\rule{0.293pt}{0.400pt}}
\multiput(575.00,337.58)(0.658,0.498){73}{\rule{0.626pt}{0.120pt}}
\multiput(575.00,336.17)(48.700,38.000){2}{\rule{0.313pt}{0.400pt}}
\multiput(625.00,375.58)(0.774,0.497){63}{\rule{0.718pt}{0.120pt}}
\multiput(625.00,374.17)(49.509,33.000){2}{\rule{0.359pt}{0.400pt}}
\multiput(676.00,408.58)(0.853,0.497){57}{\rule{0.780pt}{0.120pt}}
\multiput(676.00,407.17)(49.381,30.000){2}{\rule{0.390pt}{0.400pt}}
\multiput(727.00,438.58)(0.930,0.497){51}{\rule{0.841pt}{0.120pt}}
\multiput(727.00,437.17)(48.255,27.000){2}{\rule{0.420pt}{0.400pt}}
\multiput(777.00,465.58)(1.069,0.496){45}{\rule{0.950pt}{0.120pt}}
\multiput(777.00,464.17)(49.028,24.000){2}{\rule{0.475pt}{0.400pt}}
\multiput(828.00,489.58)(1.168,0.496){41}{\rule{1.027pt}{0.120pt}}
\multiput(828.00,488.17)(48.868,22.000){2}{\rule{0.514pt}{0.400pt}}
\multiput(879.00,511.58)(1.262,0.496){37}{\rule{1.100pt}{0.119pt}}
\multiput(879.00,510.17)(47.717,20.000){2}{\rule{0.550pt}{0.400pt}}
\multiput(929.00,531.58)(1.357,0.495){35}{\rule{1.174pt}{0.119pt}}
\multiput(929.00,530.17)(48.564,19.000){2}{\rule{0.587pt}{0.400pt}}
\multiput(980.00,550.58)(1.520,0.495){31}{\rule{1.300pt}{0.119pt}}
\multiput(980.00,549.17)(48.302,17.000){2}{\rule{0.650pt}{0.400pt}}
\multiput(1031.00,567.58)(1.586,0.494){29}{\rule{1.350pt}{0.119pt}}
\multiput(1031.00,566.17)(47.198,16.000){2}{\rule{0.675pt}{0.400pt}}
\multiput(1081.00,583.58)(1.856,0.494){25}{\rule{1.557pt}{0.119pt}}
\multiput(1081.00,582.17)(47.768,14.000){2}{\rule{0.779pt}{0.400pt}}
\multiput(1132.00,597.58)(1.856,0.494){25}{\rule{1.557pt}{0.119pt}}
\multiput(1132.00,596.17)(47.768,14.000){2}{\rule{0.779pt}{0.400pt}}
\multiput(1183.00,611.58)(2.133,0.492){21}{\rule{1.767pt}{0.119pt}}
\multiput(1183.00,610.17)(46.333,12.000){2}{\rule{0.883pt}{0.400pt}}
\multiput(1233.00,623.58)(2.176,0.492){21}{\rule{1.800pt}{0.119pt}}
\multiput(1233.00,622.17)(47.264,12.000){2}{\rule{0.900pt}{0.400pt}}
\multiput(1284.00,635.58)(2.383,0.492){19}{\rule{1.955pt}{0.118pt}}
\multiput(1284.00,634.17)(46.943,11.000){2}{\rule{0.977pt}{0.400pt}}
\multiput(1335.00,646.58)(2.580,0.491){17}{\rule{2.100pt}{0.118pt}}
\multiput(1335.00,645.17)(45.641,10.000){2}{\rule{1.050pt}{0.400pt}}
\multiput(1385.00,656.58)(2.632,0.491){17}{\rule{2.140pt}{0.118pt}}
\multiput(1385.00,655.17)(46.558,10.000){2}{\rule{1.070pt}{0.400pt}}
\put(1436,666){\usebox{\plotpoint}}
\end{picture}

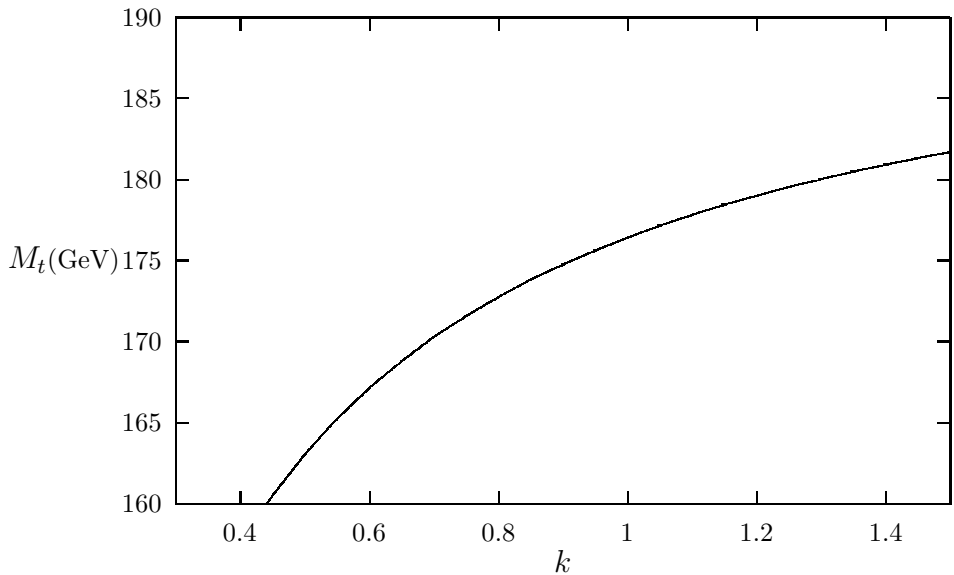
Fig. 1: The top mass $M_t$ for $k$.
\end{center}

Similarly we can calculate the bottom quark mass at the tree level 
for each value of $k$.
However, SUSY corrections to the bottom quark mass is sizable in the 
large $\tan \beta$ scenario \cite{hall} and 
that leads to another constraint \cite{COPW,Nuniver}.
Thus, we will estimate the predicted bottom quark mass with SUSY
corrections
after calculations of the SUSY mass spectrum.

We determine the values of $\mu$ and $B$-parameters by using the following 
two minimization conditions of the Higgs potential at the weak scale,
\begin{eqnarray}
&~& m_1^2+m_2^2 = -{2 \mu B \over \sin 2 \beta}, 
\label{mini1}\\
&~& m_1^2-m_2^2 = -\cos 2 \beta (M_Z^2+m_1^2+m_2^2) 
\label{mini2}
\end{eqnarray}
where $m_{1,2}^2=m_{H_1,H_2}^2+\mu^2$.

Soft SUSY breaking parameters can be constrained from requirements.
One of the most important constraints is the realization of electroweak 
symmetry breaking.
To this end, the following condition should be satisfied,
\begin{eqnarray}
m_1^2 m_2^2 < |\mu B|^2.
\label{SB}
\end{eqnarray}
In addition, the bounded-from-below condition along 
the $D$-flat direction in the Higgs potential requires 
\begin{eqnarray}
m_1^2+m_2^2 >2 |\mu B|.
\label{BFB}
\end{eqnarray}
Another important condition is the positivity of physical scalar mass
squared \footnote{It would be necessary to calculate the decay rate to the 
unbounded-from-below direction, e.g. corresponding to 
$\hat{m}_{\tau}^{2} <0$ in order to exclude completely such 
parameter region \cite{meta}.}.
For example, two stau masses squared,  
$m_{\tilde \tau_1}^2$ and $m_{\tilde \tau_2}^2$ are obtained 
as eigenvalues of the following (mass)$^2$ matrix:
\begin{equation}
\left( \begin{array}{cc}
m_{\tilde \tau_L}^2  +M_Z^2\cos2\beta(-{1 \over 2}+\sin^2\theta_W) 
& vY_\tau(A_\tau \cos \beta-\mu \sin \beta) \\
vY_\tau(A_\tau \cos \beta -\mu \sin \beta) 
& m_{\tilde \tau_R}^2 -M_Z^2\cos2\beta\sin^2\theta_W
\end{array}\right)
\end{equation}
where $v^2 \equiv\langle H_2\rangle^2 +\langle H_1\rangle ^2$.
Here we neglect the SUSY stau mass squared $M_{\tau}^2$.

Actually, in Ref. \cite{softGYU2}, it is shown these conditions 
constrain severely 
the parameter space in $SU(5)$ models with large $\tan \beta$.
Because, in the large $\tan \beta$ scenario, the stau mass squared 
receives as sizable negative corrections due to the large tau Yukawa 
coupling as the soft Higgs masses squared $m_{H_1}^2$ and $m_{H_2}^2$ do.
Large values of $m_{\tilde \tau_L}^2$ and $m_{\tilde \tau_R}^2$ at $M_X$ 
are favorable to avoid $m_{\tilde \tau_{1,2}}^2 < 0$.
Here $m_{\tilde \tau_{1}}$ denotes the lightest mass of them.
For example, it is impossible to satisfy these conditions in 
explicit $SU(5)$ models in a small value of $M$.
It is shown that the case with a common soft scalar mass, $m_i^2=M^2/3$, 
is not allowed in some finite $SU(5)$ models.

Now we discuss these constraints in generic $SO(10)$ model.
Fig. 2 shows excluded regions by these constraints for $k =1.0$
$(\tan \beta = 53)$.
In this figure, the dotted region in the left side denotes the region 
forbidden by the electroweak breaking conditions.
On the other hand, the place with asterisks correspond to the region with 
$m_{\tilde \tau_1}^2 < 0$ 
and squares denote the region where 
the light stau mass squared $m_{\tilde \tau_1}^2$ is 
smaller than the lightest neutralino mass squared $m_{\chi^0_1}^2$.
Note that, in the case with the initial condition 
$m_{\tilde \tau_L}^2=m_{\tilde \tau_R}^2$, the lightest stau ${\tilde
\tau_1}$
almost originates in ${\tilde \tau_R}$.
Because the mass squared $m_{\tilde \tau_L}^2$ has sizable positive 
radiative corrections due to $SU(2)$ gauginos, and a half size of negative
contribution from $\tau$ Yukawa coupling 
compared with that to $m_{\tilde \tau_R}^2$.
In the whole parameter space, sbottom and stop are heavier than the 
lightest stau.
\begin{center}
\setlength{\unitlength}{0.240900pt}
\ifx\plotpoint\undefined\newsavebox{\plotpoint}\fi
\sbox{\plotpoint}{\rule[-0.200pt]{0.400pt}{0.400pt}}%
\begin{picture}(1500,900)(0,0)
\font\gnuplot=cmr10 at 10pt
\gnuplot
\sbox{\plotpoint}{\rule[-0.200pt]{0.400pt}{0.400pt}}%
\put(220.0,153.0){\rule[-0.200pt]{4.818pt}{0.400pt}}
\put(198,153){\makebox(0,0)[r]{0.2}}
\put(1416.0,153.0){\rule[-0.200pt]{4.818pt}{0.400pt}}
\put(220.0,234.0){\rule[-0.200pt]{4.818pt}{0.400pt}}
\put(198,234){\makebox(0,0)[r]{0.4}}
\put(1416.0,234.0){\rule[-0.200pt]{4.818pt}{0.400pt}}
\put(220.0,314.0){\rule[-0.200pt]{4.818pt}{0.400pt}}
\put(198,314){\makebox(0,0)[r]{0.6}}
\put(1416.0,314.0){\rule[-0.200pt]{4.818pt}{0.400pt}}
\put(220.0,394.0){\rule[-0.200pt]{4.818pt}{0.400pt}}
\put(198,394){\makebox(0,0)[r]{0.8}}
\put(1416.0,394.0){\rule[-0.200pt]{4.818pt}{0.400pt}}
\put(220.0,475.0){\rule[-0.200pt]{4.818pt}{0.400pt}}
\put(198,475){\makebox(0,0)[r]{1}}
\put(1416.0,475.0){\rule[-0.200pt]{4.818pt}{0.400pt}}
\put(220.0,555.0){\rule[-0.200pt]{4.818pt}{0.400pt}}
\put(198,555){\makebox(0,0)[r]{1.2}}
\put(1416.0,555.0){\rule[-0.200pt]{4.818pt}{0.400pt}}
\put(220.0,636.0){\rule[-0.200pt]{4.818pt}{0.400pt}}
\put(198,636){\makebox(0,0)[r]{1.4}}
\put(1416.0,636.0){\rule[-0.200pt]{4.818pt}{0.400pt}}
\put(220.0,716.0){\rule[-0.200pt]{4.818pt}{0.400pt}}
\put(198,716){\makebox(0,0)[r]{1.6}}
\put(1416.0,716.0){\rule[-0.200pt]{4.818pt}{0.400pt}}
\put(220.0,797.0){\rule[-0.200pt]{4.818pt}{0.400pt}}
\put(198,797){\makebox(0,0)[r]{1.8}}
\put(1416.0,797.0){\rule[-0.200pt]{4.818pt}{0.400pt}}
\put(220.0,877.0){\rule[-0.200pt]{4.818pt}{0.400pt}}
\put(198,877){\makebox(0,0)[r]{2}}
\put(1416.0,877.0){\rule[-0.200pt]{4.818pt}{0.400pt}}
\put(284.0,113.0){\rule[-0.200pt]{0.400pt}{4.818pt}}
\put(284,68){\makebox(0,0){0.2}}
\put(284.0,857.0){\rule[-0.200pt]{0.400pt}{4.818pt}}
\put(412.0,113.0){\rule[-0.200pt]{0.400pt}{4.818pt}}
\put(412,68){\makebox(0,0){0.4}}
\put(412.0,857.0){\rule[-0.200pt]{0.400pt}{4.818pt}}
\put(540.0,113.0){\rule[-0.200pt]{0.400pt}{4.818pt}}
\put(540,68){\makebox(0,0){0.6}}
\put(540.0,857.0){\rule[-0.200pt]{0.400pt}{4.818pt}}
\put(668.0,113.0){\rule[-0.200pt]{0.400pt}{4.818pt}}
\put(668,68){\makebox(0,0){0.8}}
\put(668.0,857.0){\rule[-0.200pt]{0.400pt}{4.818pt}}
\put(796.0,113.0){\rule[-0.200pt]{0.400pt}{4.818pt}}
\put(796,68){\makebox(0,0){1}}
\put(796.0,857.0){\rule[-0.200pt]{0.400pt}{4.818pt}}
\put(924.0,113.0){\rule[-0.200pt]{0.400pt}{4.818pt}}
\put(924,68){\makebox(0,0){1.2}}
\put(924.0,857.0){\rule[-0.200pt]{0.400pt}{4.818pt}}
\put(1052.0,113.0){\rule[-0.200pt]{0.400pt}{4.818pt}}
\put(1052,68){\makebox(0,0){1.4}}
\put(1052.0,857.0){\rule[-0.200pt]{0.400pt}{4.818pt}}
\put(1180.0,113.0){\rule[-0.200pt]{0.400pt}{4.818pt}}
\put(1180,68){\makebox(0,0){1.6}}
\put(1180.0,857.0){\rule[-0.200pt]{0.400pt}{4.818pt}}
\put(1308.0,113.0){\rule[-0.200pt]{0.400pt}{4.818pt}}
\put(1308,68){\makebox(0,0){1.8}}
\put(1308.0,857.0){\rule[-0.200pt]{0.400pt}{4.818pt}}
\put(1436.0,113.0){\rule[-0.200pt]{0.400pt}{4.818pt}}
\put(1436,68){\makebox(0,0){2}}
\put(1436.0,857.0){\rule[-0.200pt]{0.400pt}{4.818pt}}
\put(220.0,113.0){\rule[-0.200pt]{292.934pt}{0.400pt}}
\put(1436.0,113.0){\rule[-0.200pt]{0.400pt}{184.048pt}}
\put(220.0,877.0){\rule[-0.200pt]{292.934pt}{0.400pt}}
\put(45,495){\makebox(0,0){$m_{16}$(TeV)}}
\put(828,23){\makebox(0,0){$M$(TeV)}}
\put(220.0,113.0){\rule[-0.200pt]{0.400pt}{184.048pt}}
\put(284,153){\circle*{12}}
\put(284,193){\circle*{12}}
\put(284,234){\circle*{12}}
\put(284,274){\circle*{12}}
\put(284,314){\circle*{12}}
\put(284,354){\circle*{12}}
\put(284,394){\circle*{12}}
\put(284,435){\circle*{12}}
\put(284,475){\circle*{12}}
\put(284,515){\circle*{12}}
\put(284,555){\circle*{12}}
\put(284,596){\circle*{12}}
\put(284,636){\circle*{12}}
\put(284,676){\circle*{12}}
\put(284,716){\circle*{12}}
\put(284,756){\circle*{12}}
\put(284,797){\circle*{12}}
\put(284,837){\circle*{12}}
\put(284,877){\circle*{12}}
\put(412,153){\makebox(0,0){$\star$}}
\put(476,153){\makebox(0,0){$\star$}}
\put(540,153){\makebox(0,0){$\star$}}
\put(604,153){\makebox(0,0){$\star$}}
\put(668,153){\makebox(0,0){$\star$}}
\put(732,153){\makebox(0,0){$\star$}}
\put(796,153){\makebox(0,0){$\star$}}
\put(860,153){\makebox(0,0){$\star$}}
\put(924,153){\makebox(0,0){$\star$}}
\put(988,153){\makebox(0,0){$\star$}}
\put(1052,153){\makebox(0,0){$\star$}}
\put(1116,153){\makebox(0,0){$\star$}}
\put(1180,153){\makebox(0,0){$\star$}}
\put(1244,153){\makebox(0,0){$\star$}}
\put(1308,153){\makebox(0,0){$\star$}}
\put(1372,153){\makebox(0,0){$\star$}}
\put(1436,153){\makebox(0,0){$\star$}}
\put(604,193){\makebox(0,0){$\star$}}
\put(668,193){\makebox(0,0){$\star$}}
\put(732,193){\makebox(0,0){$\star$}}
\put(796,193){\makebox(0,0){$\star$}}
\put(860,193){\makebox(0,0){$\star$}}
\put(924,193){\makebox(0,0){$\star$}}
\put(988,193){\makebox(0,0){$\star$}}
\put(1052,193){\makebox(0,0){$\star$}}
\put(1116,193){\makebox(0,0){$\star$}}
\put(1180,193){\makebox(0,0){$\star$}}
\put(1244,193){\makebox(0,0){$\star$}}
\put(1308,193){\makebox(0,0){$\star$}}
\put(1372,193){\makebox(0,0){$\star$}}
\put(1436,193){\makebox(0,0){$\star$}}
\put(796,234){\makebox(0,0){$\star$}}
\put(860,234){\makebox(0,0){$\star$}}
\put(924,234){\makebox(0,0){$\star$}}
\put(988,234){\makebox(0,0){$\star$}}
\put(1052,234){\makebox(0,0){$\star$}}
\put(1116,234){\makebox(0,0){$\star$}}
\put(1180,234){\makebox(0,0){$\star$}}
\put(1244,234){\makebox(0,0){$\star$}}
\put(1308,234){\makebox(0,0){$\star$}}
\put(1372,234){\makebox(0,0){$\star$}}
\put(1436,234){\makebox(0,0){$\star$}}
\put(924,274){\makebox(0,0){$\star$}}
\put(988,274){\makebox(0,0){$\star$}}
\put(1052,274){\makebox(0,0){$\star$}}
\put(1116,274){\makebox(0,0){$\star$}}
\put(1180,274){\makebox(0,0){$\star$}}
\put(1244,274){\makebox(0,0){$\star$}}
\put(1308,274){\makebox(0,0){$\star$}}
\put(1372,274){\makebox(0,0){$\star$}}
\put(1436,274){\makebox(0,0){$\star$}}
\put(1116,314){\makebox(0,0){$\star$}}
\put(1180,314){\makebox(0,0){$\star$}}
\put(1244,314){\makebox(0,0){$\star$}}
\put(1308,314){\makebox(0,0){$\star$}}
\put(1372,314){\makebox(0,0){$\star$}}
\put(1436,314){\makebox(0,0){$\star$}}
\put(1244,354){\makebox(0,0){$\star$}}
\put(1308,354){\makebox(0,0){$\star$}}
\put(1372,354){\makebox(0,0){$\star$}}
\put(1436,354){\makebox(0,0){$\star$}}
\put(1436,394){\makebox(0,0){$\star$}}
\sbox{\plotpoint}{\rule[-0.400pt]{0.800pt}{0.800pt}}%
\put(348,153){\raisebox{-.8pt}{\makebox(0,0){$\Box$}}}
\put(412,153){\raisebox{-.8pt}{\makebox(0,0){$\Box$}}}
\put(476,153){\raisebox{-.8pt}{\makebox(0,0){$\Box$}}}
\put(540,153){\raisebox{-.8pt}{\makebox(0,0){$\Box$}}}
\put(604,153){\raisebox{-.8pt}{\makebox(0,0){$\Box$}}}
\put(668,153){\raisebox{-.8pt}{\makebox(0,0){$\Box$}}}
\put(732,153){\raisebox{-.8pt}{\makebox(0,0){$\Box$}}}
\put(796,153){\raisebox{-.8pt}{\makebox(0,0){$\Box$}}}
\put(860,153){\raisebox{-.8pt}{\makebox(0,0){$\Box$}}}
\put(924,153){\raisebox{-.8pt}{\makebox(0,0){$\Box$}}}
\put(988,153){\raisebox{-.8pt}{\makebox(0,0){$\Box$}}}
\put(1052,153){\raisebox{-.8pt}{\makebox(0,0){$\Box$}}}
\put(1116,153){\raisebox{-.8pt}{\makebox(0,0){$\Box$}}}
\put(1180,153){\raisebox{-.8pt}{\makebox(0,0){$\Box$}}}
\put(1244,153){\raisebox{-.8pt}{\makebox(0,0){$\Box$}}}
\put(1308,153){\raisebox{-.8pt}{\makebox(0,0){$\Box$}}}
\put(1372,153){\raisebox{-.8pt}{\makebox(0,0){$\Box$}}}
\put(1436,153){\raisebox{-.8pt}{\makebox(0,0){$\Box$}}}
\put(476,193){\raisebox{-.8pt}{\makebox(0,0){$\Box$}}}
\put(540,193){\raisebox{-.8pt}{\makebox(0,0){$\Box$}}}
\put(604,193){\raisebox{-.8pt}{\makebox(0,0){$\Box$}}}
\put(668,193){\raisebox{-.8pt}{\makebox(0,0){$\Box$}}}
\put(732,193){\raisebox{-.8pt}{\makebox(0,0){$\Box$}}}
\put(796,193){\raisebox{-.8pt}{\makebox(0,0){$\Box$}}}
\put(860,193){\raisebox{-.8pt}{\makebox(0,0){$\Box$}}}
\put(924,193){\raisebox{-.8pt}{\makebox(0,0){$\Box$}}}
\put(988,193){\raisebox{-.8pt}{\makebox(0,0){$\Box$}}}
\put(1052,193){\raisebox{-.8pt}{\makebox(0,0){$\Box$}}}
\put(1116,193){\raisebox{-.8pt}{\makebox(0,0){$\Box$}}}
\put(1180,193){\raisebox{-.8pt}{\makebox(0,0){$\Box$}}}
\put(1244,193){\raisebox{-.8pt}{\makebox(0,0){$\Box$}}}
\put(1308,193){\raisebox{-.8pt}{\makebox(0,0){$\Box$}}}
\put(1372,193){\raisebox{-.8pt}{\makebox(0,0){$\Box$}}}
\put(1436,193){\raisebox{-.8pt}{\makebox(0,0){$\Box$}}}
\put(604,234){\raisebox{-.8pt}{\makebox(0,0){$\Box$}}}
\put(668,234){\raisebox{-.8pt}{\makebox(0,0){$\Box$}}}
\put(732,234){\raisebox{-.8pt}{\makebox(0,0){$\Box$}}}
\put(796,234){\raisebox{-.8pt}{\makebox(0,0){$\Box$}}}
\put(860,234){\raisebox{-.8pt}{\makebox(0,0){$\Box$}}}
\put(924,234){\raisebox{-.8pt}{\makebox(0,0){$\Box$}}}
\put(988,234){\raisebox{-.8pt}{\makebox(0,0){$\Box$}}}
\put(1052,234){\raisebox{-.8pt}{\makebox(0,0){$\Box$}}}
\put(1116,234){\raisebox{-.8pt}{\makebox(0,0){$\Box$}}}
\put(1180,234){\raisebox{-.8pt}{\makebox(0,0){$\Box$}}}
\put(1244,234){\raisebox{-.8pt}{\makebox(0,0){$\Box$}}}
\put(1308,234){\raisebox{-.8pt}{\makebox(0,0){$\Box$}}}
\put(1372,234){\raisebox{-.8pt}{\makebox(0,0){$\Box$}}}
\put(1436,234){\raisebox{-.8pt}{\makebox(0,0){$\Box$}}}
\put(668,274){\raisebox{-.8pt}{\makebox(0,0){$\Box$}}}
\put(732,274){\raisebox{-.8pt}{\makebox(0,0){$\Box$}}}
\put(796,274){\raisebox{-.8pt}{\makebox(0,0){$\Box$}}}
\put(860,274){\raisebox{-.8pt}{\makebox(0,0){$\Box$}}}
\put(924,274){\raisebox{-.8pt}{\makebox(0,0){$\Box$}}}
\put(988,274){\raisebox{-.8pt}{\makebox(0,0){$\Box$}}}
\put(1052,274){\raisebox{-.8pt}{\makebox(0,0){$\Box$}}}
\put(1116,274){\raisebox{-.8pt}{\makebox(0,0){$\Box$}}}
\put(1180,274){\raisebox{-.8pt}{\makebox(0,0){$\Box$}}}
\put(1244,274){\raisebox{-.8pt}{\makebox(0,0){$\Box$}}}
\put(1308,274){\raisebox{-.8pt}{\makebox(0,0){$\Box$}}}
\put(1372,274){\raisebox{-.8pt}{\makebox(0,0){$\Box$}}}
\put(1436,274){\raisebox{-.8pt}{\makebox(0,0){$\Box$}}}
\put(796,314){\raisebox{-.8pt}{\makebox(0,0){$\Box$}}}
\put(860,314){\raisebox{-.8pt}{\makebox(0,0){$\Box$}}}
\put(924,314){\raisebox{-.8pt}{\makebox(0,0){$\Box$}}}
\put(988,314){\raisebox{-.8pt}{\makebox(0,0){$\Box$}}}
\put(1052,314){\raisebox{-.8pt}{\makebox(0,0){$\Box$}}}
\put(1116,314){\raisebox{-.8pt}{\makebox(0,0){$\Box$}}}
\put(1180,314){\raisebox{-.8pt}{\makebox(0,0){$\Box$}}}
\put(1244,314){\raisebox{-.8pt}{\makebox(0,0){$\Box$}}}
\put(1308,314){\raisebox{-.8pt}{\makebox(0,0){$\Box$}}}
\put(1372,314){\raisebox{-.8pt}{\makebox(0,0){$\Box$}}}
\put(1436,314){\raisebox{-.8pt}{\makebox(0,0){$\Box$}}}
\put(924,354){\raisebox{-.8pt}{\makebox(0,0){$\Box$}}}
\put(988,354){\raisebox{-.8pt}{\makebox(0,0){$\Box$}}}
\put(1052,354){\raisebox{-.8pt}{\makebox(0,0){$\Box$}}}
\put(1116,354){\raisebox{-.8pt}{\makebox(0,0){$\Box$}}}
\put(1180,354){\raisebox{-.8pt}{\makebox(0,0){$\Box$}}}
\put(1244,354){\raisebox{-.8pt}{\makebox(0,0){$\Box$}}}
\put(1308,354){\raisebox{-.8pt}{\makebox(0,0){$\Box$}}}
\put(1372,354){\raisebox{-.8pt}{\makebox(0,0){$\Box$}}}
\put(1436,354){\raisebox{-.8pt}{\makebox(0,0){$\Box$}}}
\put(988,394){\raisebox{-.8pt}{\makebox(0,0){$\Box$}}}
\put(1052,394){\raisebox{-.8pt}{\makebox(0,0){$\Box$}}}
\put(1116,394){\raisebox{-.8pt}{\makebox(0,0){$\Box$}}}
\put(1180,394){\raisebox{-.8pt}{\makebox(0,0){$\Box$}}}
\put(1244,394){\raisebox{-.8pt}{\makebox(0,0){$\Box$}}}
\put(1308,394){\raisebox{-.8pt}{\makebox(0,0){$\Box$}}}
\put(1372,394){\raisebox{-.8pt}{\makebox(0,0){$\Box$}}}
\put(1436,394){\raisebox{-.8pt}{\makebox(0,0){$\Box$}}}
\put(1116,435){\raisebox{-.8pt}{\makebox(0,0){$\Box$}}}
\put(1180,435){\raisebox{-.8pt}{\makebox(0,0){$\Box$}}}
\put(1244,435){\raisebox{-.8pt}{\makebox(0,0){$\Box$}}}
\put(1308,435){\raisebox{-.8pt}{\makebox(0,0){$\Box$}}}
\put(1372,435){\raisebox{-.8pt}{\makebox(0,0){$\Box$}}}
\put(1436,435){\raisebox{-.8pt}{\makebox(0,0){$\Box$}}}
\put(1244,475){\raisebox{-.8pt}{\makebox(0,0){$\Box$}}}
\put(1308,475){\raisebox{-.8pt}{\makebox(0,0){$\Box$}}}
\put(1372,475){\raisebox{-.8pt}{\makebox(0,0){$\Box$}}}
\put(1436,475){\raisebox{-.8pt}{\makebox(0,0){$\Box$}}}
\put(1308,515){\raisebox{-.8pt}{\makebox(0,0){$\Box$}}}
\put(1372,515){\raisebox{-.8pt}{\makebox(0,0){$\Box$}}}
\put(1436,515){\raisebox{-.8pt}{\makebox(0,0){$\Box$}}}
\put(1436,555){\raisebox{-.8pt}{\makebox(0,0){$\Box$}}}
\end{picture}

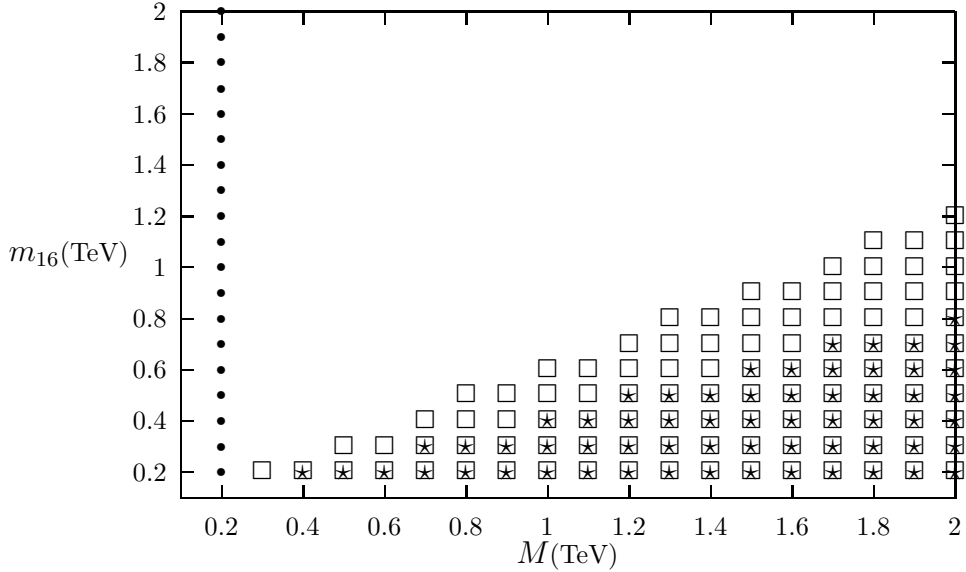
Fig.2: The allowed region by the electroweak breaking condition 
and the constraint $m_{\tilde \tau}^2 >0$ for $k=1.0$.
\end{center}

In Fig. 2, we find $m_{\tilde \tau_1}^2 < 0$ for $m_{16}< 0.4 M$
and $m_{\tilde \tau_1}^2 < m_{\chi^0_1}^2$ for 
$m_{16}< 0.6 M$.
In the universal case with $m_{16}^2=m_H^2=M^2/3$, the lightest
superpartner 
(LSP) is the lightest stau.
For other values of $k$, we obtain similar results.

In the open region of Fig. 2, the mass of the lightest 
neutral CP-even Higgs boson is 90 GeV and the 
lightest neutralino and the other superpartners as well as 
the other Higgs fields are heavier than 170 GeV.

Next we estimate the bottom quark mass at the weak scale.
The present experimental value of the bottom mass 
includes uncertainties.
For example, in Ref. \cite{bmass1}, it is shown 
\begin{equation}
m_b(M_Z)= 2.67 \pm 0.50 \ {\rm GeV}.
\label{mbex1}
\end{equation}
On the other hand, the analysis of the $\Upsilon$ system \cite{bmass2} 
and the lattice result \cite{bmass3} give $m_b(m_b) = 4.13 \pm 0.06$ GeV 
and $4.15 \pm 0.20$ GeV, respectively, \footnote{See also 
Ref.\cite{bmass4}.} which translate into
\begin{equation}
m_b(M_Z) \approx 2.8 \pm 0.2 \ {\rm GeV}.
\label{mbex2}
\end{equation}

For example, the case with $k=0.7$ in our model predicts 
$m_b(M_Z)=3.4$ GeV at the tree level.
However, the large $\tan \beta$ scenario, in general, leads to 
large SUSY corrections, i.e, 
$m_b=\lambda_b \langle H_1\rangle (1+\Delta_b)$.
Dominant contributions to $\Delta_b$ are given \cite{hall}
\begin{eqnarray}
\Delta_b&=&\frac{2 \alpha_3}{3\pi} M_{\tilde{g}} \mu \tan \beta\
 I(m_{\tilde{b}_1}^2,m_{\tilde{b}_2}^2, M_{\tilde{g}}^2)
\nonumber
\\
        &+& {Y_t^2 \over 16\pi^2} A_t \mu \tan \beta\ 
 I(m_{\tilde{t}_1}^2,m_{\tilde{t}_2}^2, \mu^2)
\label{bcor}
\end{eqnarray} 
where $M_{\tilde{g}}$, $m_{\tilde{b}_i}$ and $m_{\tilde{t}_i}$ are the
gluino, sbottom and stop eigenstate masses, respectively. 
The integral function $I(a,b,c)$ is given by
\begin{equation}
I(a,b,c)= \frac{ab\ln(a/b) +bc \ln(b/c) + ac \ln (c/a)}{(a-b)(b-c)(a-c)}.
\end{equation} 
The function $I(a,b,c)$ is of order $1/m_{max}^2$ where $m_{max}$
is the largest mass in the particles running in the corresponding loop. 
The first term of R.H.S. in eq.(\ref{bcor}) is expected to be sizable.
Since the tree level predicted value, $m_b=3.4$ GeV, 
is larger than the values given in (\ref{mbex1}) 
and (\ref{mbex2}), SUSY corrections should be negative.
That corresponds to $\mu <0$.
This region is also favorable for the constraint 
due to the $b \rightarrow s \gamma$ decay 
because this region, $\mu < 0$, can lead to smaller branching ratio 
in the large $\tan \beta$ scenario than 
the prediction by the SM \cite{bs}.
Thus we consider only the case with $\mu < 0$.
Fig. 3 shows prediction of $m_b(M_Z)$ including the correction 
$\Delta_b$ for $k=0.7$.
The curves in the figure correspond to $m_b(M_Z)=2.1$ GeV and 2.6 
GeV, which are lower bounds given in $(\ref{mbex1})$ and $(\ref{mbex2})$, 
respectively.
We have small SUSY corrections $|\Delta_b|$ in two regions, where 
$M$ is larger than $m_{16}$ and $m_{16}$ is much larger than $M$.
These regions for $\mu < 0 $ lead to the large bottom quark mass, 
e.g. $m_b(M_Z) \geq 2.6$ GeV.
This behavior is easy to see since $|\Delta_b|$ is suppressed 
when $M \gg m_{\tilde b}$ or $M \ll m_{\tilde b}$ due to the factor  
$I(m_{\tilde{b}_1}^2,m_{\tilde{b}_2}^2,M_{\tilde{g}}^2)$.
Also dotted lines in the figure show the boundaries for 
$m_{\tilde \tau_1}^2 \leq 0$ and $m_{\tilde \tau_1}^2 \leq m_{\chi
^0_1}^2$, 
which are almost same as those in Fig. 2.
The constraint $m_{\tilde \tau_1}^2 \leq 0$ excludes the region with 
$m_b(M_Z) \geq 2.6$ GeV for $m_{16} < M < 1.5$ TeV.
Furthermore, the stau is the LSP in the region with 
$m_b(M_Z) \geq 2.6$ GeV for $1.5 {\rm TeV} < m_{16} < M < 3$ TeV.
To realize $m_b(M_Z) \geq 2.6$ GeV and the neutral LSP, it is needed that 
$M > 3$ TeV or $M \ll m_{16}$ \footnote{For more precise prediction of 
the bottom mass, it is necessary to take into accuont SUSY threshold 
corrections \cite{thres}.
For example, the quasi fixed point of the bottom Yukawa coupling 
as well as the top coupling is raised due to SUSY threshold effects 
in most of cases \cite{KY}.}.
The region with $M \ll m_{16}$ can be more constrained 
by the requirement
that the LSP should not overclose the universe, i.e., 
$\Omega_{\chi} h^2 \geq 1$, 
in the case where sfermions of first and second families are 
also much heavier than the gauginos \cite{KNOY}.  
A bigger value of $k$, e.g. $k=1.0$ or 1.4, leads to larger SUSY
corrections 
$|\Delta_b|$ and predicts smaller values of the bottom mass for $\mu < 0$. 
\begin{center}
\setlength{\unitlength}{0.240900pt}
\ifx\plotpoint\undefined\newsavebox{\plotpoint}\fi
\sbox{\plotpoint}{\rule[-0.200pt]{0.400pt}{0.400pt}}%
\begin{picture}(1500,900)(0,0)
\font\gnuplot=cmr10 at 10pt
\gnuplot
\sbox{\plotpoint}{\rule[-0.200pt]{0.400pt}{0.400pt}}%
\put(220.0,161.0){\rule[-0.200pt]{4.818pt}{0.400pt}}
\put(198,161){\makebox(0,0)[r]{0.5}}
\put(1416.0,161.0){\rule[-0.200pt]{4.818pt}{0.400pt}}
\put(220.0,240.0){\rule[-0.200pt]{4.818pt}{0.400pt}}
\put(198,240){\makebox(0,0)[r]{1}}
\put(1416.0,240.0){\rule[-0.200pt]{4.818pt}{0.400pt}}
\put(220.0,320.0){\rule[-0.200pt]{4.818pt}{0.400pt}}
\put(198,320){\makebox(0,0)[r]{1.5}}
\put(1416.0,320.0){\rule[-0.200pt]{4.818pt}{0.400pt}}
\put(220.0,400.0){\rule[-0.200pt]{4.818pt}{0.400pt}}
\put(198,400){\makebox(0,0)[r]{2}}
\put(1416.0,400.0){\rule[-0.200pt]{4.818pt}{0.400pt}}
\put(220.0,479.0){\rule[-0.200pt]{4.818pt}{0.400pt}}
\put(198,479){\makebox(0,0)[r]{2.5}}
\put(1416.0,479.0){\rule[-0.200pt]{4.818pt}{0.400pt}}
\put(220.0,559.0){\rule[-0.200pt]{4.818pt}{0.400pt}}
\put(198,559){\makebox(0,0)[r]{3}}
\put(1416.0,559.0){\rule[-0.200pt]{4.818pt}{0.400pt}}
\put(220.0,638.0){\rule[-0.200pt]{4.818pt}{0.400pt}}
\put(198,638){\makebox(0,0)[r]{3.5}}
\put(1416.0,638.0){\rule[-0.200pt]{4.818pt}{0.400pt}}
\put(220.0,718.0){\rule[-0.200pt]{4.818pt}{0.400pt}}
\put(198,718){\makebox(0,0)[r]{4}}
\put(1416.0,718.0){\rule[-0.200pt]{4.818pt}{0.400pt}}
\put(220.0,797.0){\rule[-0.200pt]{4.818pt}{0.400pt}}
\put(198,797){\makebox(0,0)[r]{4.5}}
\put(1416.0,797.0){\rule[-0.200pt]{4.818pt}{0.400pt}}
\put(220.0,877.0){\rule[-0.200pt]{4.818pt}{0.400pt}}
\put(198,877){\makebox(0,0)[r]{5}}
\put(1416.0,877.0){\rule[-0.200pt]{4.818pt}{0.400pt}}
\put(296.0,113.0){\rule[-0.200pt]{0.400pt}{4.818pt}}
\put(296,68){\makebox(0,0){0.5}}
\put(296.0,857.0){\rule[-0.200pt]{0.400pt}{4.818pt}}
\put(423.0,113.0){\rule[-0.200pt]{0.400pt}{4.818pt}}
\put(423,68){\makebox(0,0){1}}
\put(423.0,857.0){\rule[-0.200pt]{0.400pt}{4.818pt}}
\put(549.0,113.0){\rule[-0.200pt]{0.400pt}{4.818pt}}
\put(549,68){\makebox(0,0){1.5}}
\put(549.0,857.0){\rule[-0.200pt]{0.400pt}{4.818pt}}
\put(676.0,113.0){\rule[-0.200pt]{0.400pt}{4.818pt}}
\put(676,68){\makebox(0,0){2}}
\put(676.0,857.0){\rule[-0.200pt]{0.400pt}{4.818pt}}
\put(803.0,113.0){\rule[-0.200pt]{0.400pt}{4.818pt}}
\put(803,68){\makebox(0,0){2.5}}
\put(803.0,857.0){\rule[-0.200pt]{0.400pt}{4.818pt}}
\put(929.0,113.0){\rule[-0.200pt]{0.400pt}{4.818pt}}
\put(929,68){\makebox(0,0){3}}
\put(929.0,857.0){\rule[-0.200pt]{0.400pt}{4.818pt}}
\put(1056.0,113.0){\rule[-0.200pt]{0.400pt}{4.818pt}}
\put(1056,68){\makebox(0,0){3.5}}
\put(1056.0,857.0){\rule[-0.200pt]{0.400pt}{4.818pt}}
\put(1183.0,113.0){\rule[-0.200pt]{0.400pt}{4.818pt}}
\put(1183,68){\makebox(0,0){4}}
\put(1183.0,857.0){\rule[-0.200pt]{0.400pt}{4.818pt}}
\put(1309.0,113.0){\rule[-0.200pt]{0.400pt}{4.818pt}}
\put(1309,68){\makebox(0,0){4.5}}
\put(1309.0,857.0){\rule[-0.200pt]{0.400pt}{4.818pt}}
\put(1436.0,113.0){\rule[-0.200pt]{0.400pt}{4.818pt}}
\put(1436,68){\makebox(0,0){5}}
\put(1436.0,857.0){\rule[-0.200pt]{0.400pt}{4.818pt}}
\put(220.0,113.0){\rule[-0.200pt]{292.934pt}{0.400pt}}
\put(1436.0,113.0){\rule[-0.200pt]{0.400pt}{184.048pt}}
\put(220.0,877.0){\rule[-0.200pt]{292.934pt}{0.400pt}}
\put(45,495){\makebox(0,0){$m_{16}$(TeV)}}
\put(828,23){\makebox(0,0){$M$(TeV)}}
\put(423,829){\makebox(0,0)[r]{2.6 GeV}}
\put(752,161){\makebox(0,0)[r]{2.6 GeV}}
\put(423,479){\makebox(0,0)[r]{2.1 GeV}}
\put(1132,829){\makebox(0,0)[r]{2.1 GeV}}
\put(1385,320){\makebox(0,0)[r]{$m_{\tau_1}^2 <0$}}
\put(1385,575){\makebox(0,0)[r]{$m_{\tau_1}^2 =m_{\chi ^0_1}^2$}}
\put(220.0,113.0){\rule[-0.200pt]{0.400pt}{184.048pt}}
\put(220,113){\usebox{\plotpoint}}
\multiput(220.00,113.58)(0.685,0.499){219}{\rule{0.648pt}{0.120pt}}
\multiput(220.00,112.17)(150.656,111.000){2}{\rule{0.324pt}{0.400pt}}
\multiput(372.00,224.58)(0.567,0.499){221}{\rule{0.554pt}{0.120pt}}
\multiput(372.00,223.17)(125.851,112.000){2}{\rule{0.277pt}{0.400pt}}
\multiput(499.00,336.58)(0.531,0.499){187}{\rule{0.525pt}{0.120pt}}
\multiput(499.00,335.17)(99.910,95.000){2}{\rule{0.263pt}{0.400pt}}
\multiput(600.58,431.00)(0.499,0.554){199}{\rule{0.120pt}{0.544pt}}
\multiput(599.17,431.00)(101.000,110.872){2}{\rule{0.400pt}{0.272pt}}
\multiput(701.58,543.00)(0.499,0.519){149}{\rule{0.120pt}{0.516pt}}
\multiput(700.17,543.00)(76.000,77.929){2}{\rule{0.400pt}{0.258pt}}
\multiput(777.58,622.00)(0.499,0.576){301}{\rule{0.120pt}{0.561pt}}
\multiput(776.17,622.00)(152.000,173.837){2}{\rule{0.400pt}{0.280pt}}
\multiput(929.58,797.00)(0.498,0.786){99}{\rule{0.120pt}{0.727pt}}
\multiput(928.17,797.00)(51.000,78.490){2}{\rule{0.400pt}{0.364pt}}
\put(220,400){\usebox{\plotpoint}}
\multiput(220.58,400.00)(0.499,1.571){301}{\rule{0.120pt}{1.355pt}}
\multiput(219.17,400.00)(152.000,474.187){2}{\rule{0.400pt}{0.678pt}}
\put(220,670){\usebox{\plotpoint}}
\multiput(220.58,670.00)(0.498,2.039){99}{\rule{0.120pt}{1.724pt}}
\multiput(219.17,670.00)(51.000,203.423){2}{\rule{0.400pt}{0.862pt}}
\put(397,113){\usebox{\plotpoint}}
\multiput(397.00,113.58)(1.041,0.500){825}{\rule{0.933pt}{0.120pt}}
\multiput(397.00,112.17)(860.064,414.000){2}{\rule{0.466pt}{0.400pt}}
\multiput(1259.00,527.58)(0.933,0.499){187}{\rule{0.845pt}{0.120pt}}
\multiput(1259.00,526.17)(175.246,95.000){2}{\rule{0.423pt}{0.400pt}}
\sbox{\plotpoint}{\rule[-0.500pt]{1.000pt}{1.000pt}}%
\put(296.00,113.00){\usebox{\plotpoint}}
\put(316.01,118.50){\usebox{\plotpoint}}
\multiput(318,119)(20.224,4.667){0}{\usebox{\plotpoint}}
\put(336.20,123.30){\usebox{\plotpoint}}
\multiput(343,125)(20.136,5.034){0}{\usebox{\plotpoint}}
\put(356.34,128.33){\usebox{\plotpoint}}
\put(376.51,133.20){\usebox{\plotpoint}}
\multiput(380,134)(20.136,5.034){0}{\usebox{\plotpoint}}
\put(396.66,138.17){\usebox{\plotpoint}}
\put(416.86,142.97){\usebox{\plotpoint}}
\multiput(417,143)(20.136,5.034){0}{\usebox{\plotpoint}}
\put(436.99,148.00){\usebox{\plotpoint}}
\multiput(441,149)(19.690,6.563){0}{\usebox{\plotpoint}}
\put(456.87,153.89){\usebox{\plotpoint}}
\put(477.05,158.76){\usebox{\plotpoint}}
\multiput(478,159)(20.136,5.034){0}{\usebox{\plotpoint}}
\put(497.22,163.67){\usebox{\plotpoint}}
\multiput(503,165)(20.136,5.034){0}{\usebox{\plotpoint}}
\put(517.38,168.59){\usebox{\plotpoint}}
\put(537.51,173.63){\usebox{\plotpoint}}
\multiput(539,174)(20.224,4.667){0}{\usebox{\plotpoint}}
\put(557.71,178.43){\usebox{\plotpoint}}
\multiput(564,180)(20.136,5.034){0}{\usebox{\plotpoint}}
\put(577.80,183.60){\usebox{\plotpoint}}
\put(597.75,189.25){\usebox{\plotpoint}}
\multiput(601,190)(20.136,5.034){0}{\usebox{\plotpoint}}
\put(617.90,194.22){\usebox{\plotpoint}}
\multiput(625,196)(20.224,4.667){0}{\usebox{\plotpoint}}
\put(638.09,199.02){\usebox{\plotpoint}}
\put(658.23,204.06){\usebox{\plotpoint}}
\multiput(662,205)(20.136,5.034){0}{\usebox{\plotpoint}}
\put(678.38,209.01){\usebox{\plotpoint}}
\put(698.55,213.89){\usebox{\plotpoint}}
\multiput(699,214)(20.136,5.034){0}{\usebox{\plotpoint}}
\put(718.72,218.78){\usebox{\plotpoint}}
\multiput(724,220)(19.690,6.563){0}{\usebox{\plotpoint}}
\put(738.61,224.65){\usebox{\plotpoint}}
\put(758.75,229.69){\usebox{\plotpoint}}
\multiput(760,230)(20.224,4.667){0}{\usebox{\plotpoint}}
\put(778.94,234.48){\usebox{\plotpoint}}
\multiput(785,236)(20.136,5.034){0}{\usebox{\plotpoint}}
\put(799.08,239.48){\usebox{\plotpoint}}
\put(819.27,244.32){\usebox{\plotpoint}}
\multiput(822,245)(20.136,5.034){0}{\usebox{\plotpoint}}
\put(839.40,249.35){\usebox{\plotpoint}}
\multiput(846,251)(20.224,4.667){0}{\usebox{\plotpoint}}
\put(859.58,254.19){\usebox{\plotpoint}}
\put(879.46,260.12){\usebox{\plotpoint}}
\multiput(883,261)(20.224,4.667){0}{\usebox{\plotpoint}}
\put(899.65,264.91){\usebox{\plotpoint}}
\put(919.79,269.95){\usebox{\plotpoint}}
\multiput(920,270)(20.136,5.034){0}{\usebox{\plotpoint}}
\put(939.96,274.84){\usebox{\plotpoint}}
\multiput(945,276)(20.136,5.034){0}{\usebox{\plotpoint}}
\put(960.12,279.78){\usebox{\plotpoint}}
\put(980.30,284.61){\usebox{\plotpoint}}
\multiput(982,285)(20.136,5.034){0}{\usebox{\plotpoint}}
\put(1000.45,289.61){\usebox{\plotpoint}}
\multiput(1006,291)(19.690,6.563){0}{\usebox{\plotpoint}}
\put(1020.32,295.54){\usebox{\plotpoint}}
\put(1040.50,300.38){\usebox{\plotpoint}}
\multiput(1043,301)(20.136,5.034){0}{\usebox{\plotpoint}}
\put(1060.66,305.31){\usebox{\plotpoint}}
\multiput(1068,307)(20.136,5.034){0}{\usebox{\plotpoint}}
\put(1080.83,310.21){\usebox{\plotpoint}}
\put(1100.97,315.24){\usebox{\plotpoint}}
\multiput(1104,316)(20.224,4.667){0}{\usebox{\plotpoint}}
\put(1121.16,320.04){\usebox{\plotpoint}}
\multiput(1129,322)(20.136,5.034){0}{\usebox{\plotpoint}}
\put(1141.29,325.10){\usebox{\plotpoint}}
\put(1161.19,330.89){\usebox{\plotpoint}}
\multiput(1166,332)(20.136,5.034){0}{\usebox{\plotpoint}}
\put(1181.35,335.84){\usebox{\plotpoint}}
\put(1201.54,340.66){\usebox{\plotpoint}}
\multiput(1203,341)(20.136,5.034){0}{\usebox{\plotpoint}}
\put(1221.68,345.67){\usebox{\plotpoint}}
\multiput(1227,347)(20.136,5.034){0}{\usebox{\plotpoint}}
\put(1241.83,350.65){\usebox{\plotpoint}}
\put(1262.01,355.50){\usebox{\plotpoint}}
\multiput(1264,356)(20.136,5.034){0}{\usebox{\plotpoint}}
\put(1282.17,360.42){\usebox{\plotpoint}}
\multiput(1289,362)(19.690,6.563){0}{\usebox{\plotpoint}}
\put(1302.07,366.27){\usebox{\plotpoint}}
\put(1322.20,371.30){\usebox{\plotpoint}}
\multiput(1325,372)(20.224,4.667){0}{\usebox{\plotpoint}}
\put(1342.39,376.10){\usebox{\plotpoint}}
\multiput(1350,378)(20.136,5.034){0}{\usebox{\plotpoint}}
\put(1362.53,381.12){\usebox{\plotpoint}}
\put(1382.72,385.93){\usebox{\plotpoint}}
\multiput(1387,387)(20.136,5.034){0}{\usebox{\plotpoint}}
\put(1402.86,390.96){\usebox{\plotpoint}}
\put(1423.05,395.78){\usebox{\plotpoint}}
\multiput(1424,396)(19.690,6.563){0}{\usebox{\plotpoint}}
\put(1436,400){\usebox{\plotpoint}}
\put(254.00,113.00){\usebox{\plotpoint}}
\multiput(257,114)(19.159,7.983){0}{\usebox{\plotpoint}}
\put(273.36,120.45){\usebox{\plotpoint}}
\put(292.85,127.56){\usebox{\plotpoint}}
\multiput(294,128)(19.159,7.983){0}{\usebox{\plotpoint}}
\put(312.19,135.06){\usebox{\plotpoint}}
\multiput(318,137)(19.372,7.451){0}{\usebox{\plotpoint}}
\put(331.65,142.27){\usebox{\plotpoint}}
\put(351.03,149.68){\usebox{\plotpoint}}
\multiput(355,151)(19.159,7.983){0}{\usebox{\plotpoint}}
\put(370.41,157.05){\usebox{\plotpoint}}
\put(389.90,164.12){\usebox{\plotpoint}}
\multiput(392,165)(19.159,7.983){0}{\usebox{\plotpoint}}
\put(409.23,171.61){\usebox{\plotpoint}}
\put(428.66,178.86){\usebox{\plotpoint}}
\multiput(429,179)(19.159,7.983){0}{\usebox{\plotpoint}}
\put(448.01,186.34){\usebox{\plotpoint}}
\multiput(453,188)(19.372,7.451){0}{\usebox{\plotpoint}}
\put(467.44,193.60){\usebox{\plotpoint}}
\put(486.84,200.95){\usebox{\plotpoint}}
\multiput(490,202)(19.372,7.451){0}{\usebox{\plotpoint}}
\put(506.32,208.11){\usebox{\plotpoint}}
\put(525.71,215.46){\usebox{\plotpoint}}
\multiput(527,216)(19.159,7.983){0}{\usebox{\plotpoint}}
\put(545.08,222.87){\usebox{\plotpoint}}
\multiput(552,225)(19.159,7.983){0}{\usebox{\plotpoint}}
\put(564.47,230.20){\usebox{\plotpoint}}
\put(583.84,237.61){\usebox{\plotpoint}}
\multiput(588,239)(19.372,7.451){0}{\usebox{\plotpoint}}
\put(603.32,244.77){\usebox{\plotpoint}}
\put(622.74,252.06){\usebox{\plotpoint}}
\multiput(625,253)(19.372,7.451){0}{\usebox{\plotpoint}}
\put(642.16,259.39){\usebox{\plotpoint}}
\put(661.53,266.80){\usebox{\plotpoint}}
\multiput(662,267)(19.159,7.983){0}{\usebox{\plotpoint}}
\put(680.92,274.13){\usebox{\plotpoint}}
\multiput(687,276)(19.159,7.983){0}{\usebox{\plotpoint}}
\put(700.32,281.44){\usebox{\plotpoint}}
\put(719.87,288.41){\usebox{\plotpoint}}
\multiput(724,290)(19.159,7.983){0}{\usebox{\plotpoint}}
\put(739.16,296.05){\usebox{\plotpoint}}
\put(758.56,303.40){\usebox{\plotpoint}}
\multiput(760,304)(19.372,7.451){0}{\usebox{\plotpoint}}
\put(777.99,310.66){\usebox{\plotpoint}}
\multiput(785,313)(19.159,7.983){0}{\usebox{\plotpoint}}
\put(797.34,318.13){\usebox{\plotpoint}}
\put(816.83,325.28){\usebox{\plotpoint}}
\multiput(822,327)(19.159,7.983){0}{\usebox{\plotpoint}}
\put(836.18,332.73){\usebox{\plotpoint}}
\put(855.72,339.74){\usebox{\plotpoint}}
\multiput(859,341)(19.159,7.983){0}{\usebox{\plotpoint}}
\put(875.02,347.34){\usebox{\plotpoint}}
\put(894.52,354.43){\usebox{\plotpoint}}
\multiput(896,355)(19.159,7.983){0}{\usebox{\plotpoint}}
\put(913.85,361.95){\usebox{\plotpoint}}
\multiput(920,364)(19.159,7.983){0}{\usebox{\plotpoint}}
\put(933.22,369.38){\usebox{\plotpoint}}
\put(952.78,376.24){\usebox{\plotpoint}}
\multiput(957,378)(19.159,7.983){0}{\usebox{\plotpoint}}
\put(972.05,383.94){\usebox{\plotpoint}}
\put(991.54,390.98){\usebox{\plotpoint}}
\multiput(994,392)(19.159,7.983){0}{\usebox{\plotpoint}}
\put(1010.83,398.61){\usebox{\plotpoint}}
\put(1030.32,405.74){\usebox{\plotpoint}}
\multiput(1031,406)(19.690,6.563){0}{\usebox{\plotpoint}}
\put(1049.81,412.84){\usebox{\plotpoint}}
\multiput(1055,415)(19.372,7.451){0}{\usebox{\plotpoint}}
\put(1069.15,420.38){\usebox{\plotpoint}}
\put(1088.60,427.58){\usebox{\plotpoint}}
\multiput(1092,429)(19.159,7.983){0}{\usebox{\plotpoint}}
\put(1107.89,435.20){\usebox{\plotpoint}}
\put(1127.36,442.32){\usebox{\plotpoint}}
\multiput(1129,443)(19.159,7.983){0}{\usebox{\plotpoint}}
\put(1146.67,449.89){\usebox{\plotpoint}}
\multiput(1153,452)(19.372,7.451){0}{\usebox{\plotpoint}}
\put(1166.15,457.05){\usebox{\plotpoint}}
\put(1185.63,464.18){\usebox{\plotpoint}}
\multiput(1190,466)(19.372,7.451){0}{\usebox{\plotpoint}}
\put(1204.98,471.66){\usebox{\plotpoint}}
\put(1224.41,478.92){\usebox{\plotpoint}}
\multiput(1227,480)(19.159,7.983){0}{\usebox{\plotpoint}}
\put(1243.73,486.46){\usebox{\plotpoint}}
\put(1263.17,493.66){\usebox{\plotpoint}}
\multiput(1264,494)(19.690,6.563){0}{\usebox{\plotpoint}}
\put(1282.73,500.59){\usebox{\plotpoint}}
\multiput(1289,503)(19.159,7.983){0}{\usebox{\plotpoint}}
\put(1301.99,508.33){\usebox{\plotpoint}}
\put(1321.44,515.52){\usebox{\plotpoint}}
\multiput(1325,517)(19.372,7.451){0}{\usebox{\plotpoint}}
\put(1340.82,522.94){\usebox{\plotpoint}}
\put(1360.23,530.26){\usebox{\plotpoint}}
\multiput(1362,531)(19.372,7.451){0}{\usebox{\plotpoint}}
\put(1379.65,537.55){\usebox{\plotpoint}}
\multiput(1387,540)(19.159,7.983){0}{\usebox{\plotpoint}}
\put(1399.01,545.00){\usebox{\plotpoint}}
\put(1418.58,551.91){\usebox{\plotpoint}}
\multiput(1424,554)(19.159,7.983){0}{\usebox{\plotpoint}}
\put(1436,559){\usebox{\plotpoint}}
\end{picture}

Fig.3: Predicted values of $m_b(M_Z)$ for $k=0.7$.
\end{center}

We have discussed the case that squarks and sleptons in the third 
family have the same soft scalar mass $m_{16}$ at $M_X$,
which is required by unbroken $SO(10)$ gauge symmetry.
However, if a gauge symmetry breaks reducing its rank like 
$SO(10) \rightarrow G_{SM}$, additional sizable 
contributions to soft scalar masses can appear at the breaking scale, 
which is called $D$-term contributions \cite{Dterm1,Dterm2}.
These $D$-term contributions are, in general, proportional to 
quantum numbers of broken diagonal generators.
If we specify the model, one can calculate 
their magnitudes \cite{Dterm3}.
Here we study the effect on parameter space keeping them free parameters.
The soft scalar masses at $M_X$ are written as
\begin{eqnarray}
&~&m_{\tilde Q}^2=m_{\tilde t}^2=m_{\tilde \tau_R}^2=m_{16}^2-m_D^2, \quad
m_{\tilde b}^2=m_{\tilde \tau_L}^2=m_{16}^2+3m_D^2, 
\nonumber\\
&~&m_{H_1}^2=M^2-2m_{16}^2-2m_D^2,\quad
m_{H_2}^2=M^2-2m_{16}^2+2m_D^2
\end{eqnarray}
in the presence of the $D$-term contributions, $Q_{10} m_D^2$, 
where $Q_{10}$ denotes a broken diagonal charge up to a normalization
factor.
Note that the above soft masses satisfy the sum 
rule (\ref{sumrule1}) even with taking into account 
$D$-term contributions.
Because $D$-term contributions are proportional to broken charges 
and these charges should conserve in allowed couplings.

A positive value of $m_D^2$ is unfavorable for successful 
electroweak breaking and the constraint $m_{\tilde \tau_1}^2 >0$
since it reduces the values of
$m_{H_1}^2$ and $m_{\tilde \tau_R}^2$.
Recall that $m_{\tilde \tau_R}$ is dominant to $m_{\tilde \tau_1}$ with
the 
initial condition $m_{\tilde \tau_L}^2=m_{\tilde \tau_R}^2$. 
For example, in the case with $m_D^2=M^2/3$, the whole parameter 
space of $(M,m_{16})$ is ruled out from the electroweak symmetry
breaking condition.
On the other hand, a negative value of $m_D^2$ is favorable to 
increase both $m_{H_1}^2$ and $m_{\tilde \tau_R}^2$,
while it reduces $m_{\tilde \tau_L}^2$.
The region excluded by the electroweak breaking condition     
easily becomes narrow for $m_D^2<0$.
The constraint due to the lightest stau is also relaxed 
as  $m_D^2$ decreases from $m_D^2=0$.
However, that reduces $m_{\tilde \tau_L}^2$ and below a critical value 
of $m_D^2$ the value of $m_{\tilde \tau_L}^2$ becomes dominant to 
$m_{\tilde \tau_1}^2$.
Thus below such a critical value of $m_D^2$, the allowed region becomes 
narrow due to the constraint for $m_{\tilde \tau_1}^2$.
Around $m_D^2/M^2=-0.1$ we obtain the widest allowed region.
Fig. 4 shows the case with $m_D^2/M^2=-0.1$ and $k=0.7$.
In this case, the constraint due to the electroweak breaking is 
less important, and actually the excluded region by the electroweak 
breaking is out of the region shown in Fig.4.
Dotted lines correspond to boundaries for 
$m_{\tilde \tau_1}^2 \leq 0$ and $m_{\tilde \tau_1}^2 \leq m_{\chi
^0_1}^2$. 
The region with $m_{\tilde \tau_1}^2 \leq 0$ becomes narrow.
The predicted values of the bottom quark mass is shown 
as curves corresponding to $2.1$ GeV and 2.6 GeV in Fig. 4.
For $M > m_{16}$, we have large SUSY corrections $|\Delta_b|$ compared 
with the $m_D^2=0$ case.
\begin{center}
\setlength{\unitlength}{0.240900pt}
\ifx\plotpoint\undefined\newsavebox{\plotpoint}\fi
\sbox{\plotpoint}{\rule[-0.200pt]{0.400pt}{0.400pt}}%
\begin{picture}(1500,900)(0,0)
\font\gnuplot=cmr10 at 10pt
\gnuplot
\sbox{\plotpoint}{\rule[-0.200pt]{0.400pt}{0.400pt}}%
\put(220.0,161.0){\rule[-0.200pt]{4.818pt}{0.400pt}}
\put(198,161){\makebox(0,0)[r]{0.5}}
\put(1416.0,161.0){\rule[-0.200pt]{4.818pt}{0.400pt}}
\put(220.0,240.0){\rule[-0.200pt]{4.818pt}{0.400pt}}
\put(198,240){\makebox(0,0)[r]{1}}
\put(1416.0,240.0){\rule[-0.200pt]{4.818pt}{0.400pt}}
\put(220.0,320.0){\rule[-0.200pt]{4.818pt}{0.400pt}}
\put(198,320){\makebox(0,0)[r]{1.5}}
\put(1416.0,320.0){\rule[-0.200pt]{4.818pt}{0.400pt}}
\put(220.0,400.0){\rule[-0.200pt]{4.818pt}{0.400pt}}
\put(198,400){\makebox(0,0)[r]{2}}
\put(1416.0,400.0){\rule[-0.200pt]{4.818pt}{0.400pt}}
\put(220.0,479.0){\rule[-0.200pt]{4.818pt}{0.400pt}}
\put(198,479){\makebox(0,0)[r]{2.5}}
\put(1416.0,479.0){\rule[-0.200pt]{4.818pt}{0.400pt}}
\put(220.0,559.0){\rule[-0.200pt]{4.818pt}{0.400pt}}
\put(198,559){\makebox(0,0)[r]{3}}
\put(1416.0,559.0){\rule[-0.200pt]{4.818pt}{0.400pt}}
\put(220.0,638.0){\rule[-0.200pt]{4.818pt}{0.400pt}}
\put(198,638){\makebox(0,0)[r]{3.5}}
\put(1416.0,638.0){\rule[-0.200pt]{4.818pt}{0.400pt}}
\put(220.0,718.0){\rule[-0.200pt]{4.818pt}{0.400pt}}
\put(198,718){\makebox(0,0)[r]{4}}
\put(1416.0,718.0){\rule[-0.200pt]{4.818pt}{0.400pt}}
\put(220.0,797.0){\rule[-0.200pt]{4.818pt}{0.400pt}}
\put(198,797){\makebox(0,0)[r]{4.5}}
\put(1416.0,797.0){\rule[-0.200pt]{4.818pt}{0.400pt}}
\put(220.0,877.0){\rule[-0.200pt]{4.818pt}{0.400pt}}
\put(198,877){\makebox(0,0)[r]{5}}
\put(1416.0,877.0){\rule[-0.200pt]{4.818pt}{0.400pt}}
\put(296.0,113.0){\rule[-0.200pt]{0.400pt}{4.818pt}}
\put(296,68){\makebox(0,0){0.5}}
\put(296.0,857.0){\rule[-0.200pt]{0.400pt}{4.818pt}}
\put(423.0,113.0){\rule[-0.200pt]{0.400pt}{4.818pt}}
\put(423,68){\makebox(0,0){1}}
\put(423.0,857.0){\rule[-0.200pt]{0.400pt}{4.818pt}}
\put(549.0,113.0){\rule[-0.200pt]{0.400pt}{4.818pt}}
\put(549,68){\makebox(0,0){1.5}}
\put(549.0,857.0){\rule[-0.200pt]{0.400pt}{4.818pt}}
\put(676.0,113.0){\rule[-0.200pt]{0.400pt}{4.818pt}}
\put(676,68){\makebox(0,0){2}}
\put(676.0,857.0){\rule[-0.200pt]{0.400pt}{4.818pt}}
\put(803.0,113.0){\rule[-0.200pt]{0.400pt}{4.818pt}}
\put(803,68){\makebox(0,0){2.5}}
\put(803.0,857.0){\rule[-0.200pt]{0.400pt}{4.818pt}}
\put(929.0,113.0){\rule[-0.200pt]{0.400pt}{4.818pt}}
\put(929,68){\makebox(0,0){3}}
\put(929.0,857.0){\rule[-0.200pt]{0.400pt}{4.818pt}}
\put(1056.0,113.0){\rule[-0.200pt]{0.400pt}{4.818pt}}
\put(1056,68){\makebox(0,0){3.5}}
\put(1056.0,857.0){\rule[-0.200pt]{0.400pt}{4.818pt}}
\put(1183.0,113.0){\rule[-0.200pt]{0.400pt}{4.818pt}}
\put(1183,68){\makebox(0,0){4}}
\put(1183.0,857.0){\rule[-0.200pt]{0.400pt}{4.818pt}}
\put(1309.0,113.0){\rule[-0.200pt]{0.400pt}{4.818pt}}
\put(1309,68){\makebox(0,0){4.5}}
\put(1309.0,857.0){\rule[-0.200pt]{0.400pt}{4.818pt}}
\put(1436.0,113.0){\rule[-0.200pt]{0.400pt}{4.818pt}}
\put(1436,68){\makebox(0,0){5}}
\put(1436.0,857.0){\rule[-0.200pt]{0.400pt}{4.818pt}}
\put(220.0,113.0){\rule[-0.200pt]{292.934pt}{0.400pt}}
\put(1436.0,113.0){\rule[-0.200pt]{0.400pt}{184.048pt}}
\put(220.0,877.0){\rule[-0.200pt]{292.934pt}{0.400pt}}
\put(45,495){\makebox(0,0){$m_{16}$(TeV)}}
\put(828,23){\makebox(0,0){$M$(TeV)}}
\put(423,829){\makebox(0,0)[r]{2.6 GeV}}
\put(1259,320){\makebox(0,0)[r]{2.6 GeV}}
\put(423,479){\makebox(0,0)[r]{2.1 GeV}}
\put(1183,829){\makebox(0,0)[r]{2.1 GeV}}
\put(1005,145){\makebox(0,0)[r]{$m_{\tau_1}^2 <0$}}
\put(1411,559){\makebox(0,0)[r]{$m_{\tau_1}^2 =m_{\chi ^0_1}^2$}}
\put(220.0,113.0){\rule[-0.200pt]{0.400pt}{184.048pt}}
\put(220,400){\usebox{\plotpoint}}
\multiput(220.58,400.00)(0.499,1.571){301}{\rule{0.120pt}{1.355pt}}
\multiput(219.17,400.00)(152.000,474.187){2}{\rule{0.400pt}{0.678pt}}
\put(245,113){\usebox{\plotpoint}}
\multiput(245.00,113.58)(0.665,0.500){379}{\rule{0.632pt}{0.120pt}}
\multiput(245.00,112.17)(252.688,191.000){2}{\rule{0.316pt}{0.400pt}}
\multiput(499.00,304.58)(0.567,0.499){219}{\rule{0.554pt}{0.120pt}}
\multiput(499.00,303.17)(124.850,111.000){2}{\rule{0.277pt}{0.400pt}}
\multiput(625.00,415.58)(0.567,0.499){221}{\rule{0.554pt}{0.120pt}}
\multiput(625.00,414.17)(125.851,112.000){2}{\rule{0.277pt}{0.400pt}}
\multiput(752.58,527.00)(0.499,0.519){149}{\rule{0.120pt}{0.516pt}}
\multiput(751.17,527.00)(76.000,77.929){2}{\rule{0.400pt}{0.258pt}}
\multiput(828.58,606.00)(0.500,0.594){453}{\rule{0.120pt}{0.575pt}}
\multiput(827.17,606.00)(228.000,269.806){2}{\rule{0.400pt}{0.288pt}}
\put(220,670){\usebox{\plotpoint}}
\multiput(220.58,670.00)(0.498,2.039){99}{\rule{0.120pt}{1.724pt}}
\multiput(219.17,670.00)(51.000,203.423){2}{\rule{0.400pt}{0.862pt}}
\put(575,113){\usebox{\plotpoint}}
\multiput(575.00,113.58)(1.177,0.500){729}{\rule{1.041pt}{0.120pt}}
\multiput(575.00,112.17)(858.839,366.000){2}{\rule{0.520pt}{0.400pt}}
\sbox{\plotpoint}{\rule[-0.500pt]{1.000pt}{1.000pt}}%
\put(245,113){\usebox{\plotpoint}}
\multiput(245,113)(20.607,2.474){58}{\usebox{\plotpoint}}
\put(1436,256){\usebox{\plotpoint}}
\put(245,113){\usebox{\plotpoint}}
\multiput(245,113)(19.764,6.339){61}{\usebox{\plotpoint}}
\put(1436,495){\usebox{\plotpoint}}
\end{picture}

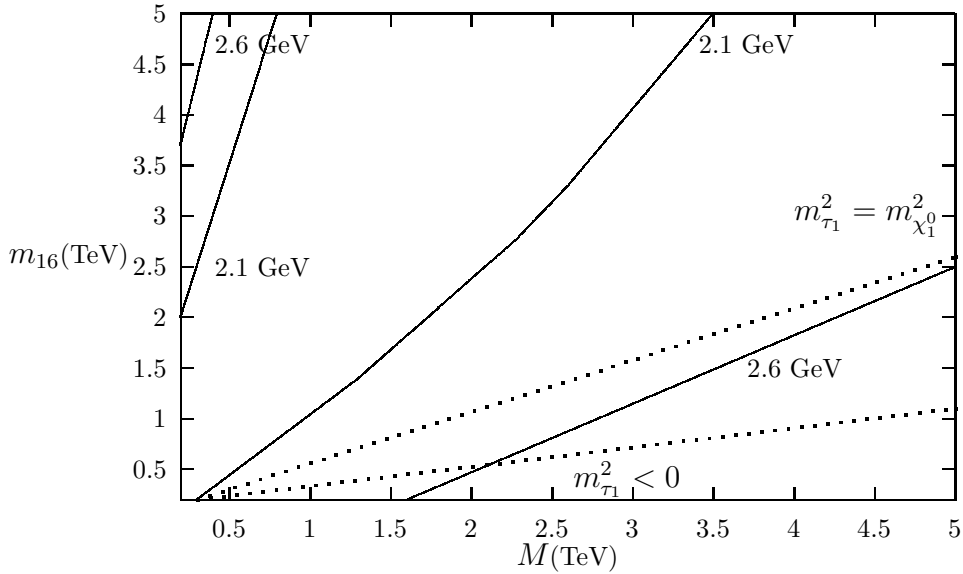
Fig.4: The excluded region by the constraint $m_{\tilde \tau}^2 > 0$ 
and predicted values of $m_b(M_Z)$ for $k=0.7$ and $m_D^2/M^2=-0.1$.
\end{center}

If $m_D^2/M^2 < -0.1$, wider region is excluded by the stau mass
constraint.
On top of that, SUSY corrections $|\Delta_b|$ become large in the region 
$M > m_{16}$, although the opposite region with $M \ll m_{16}$ leads 
to slightly small SUSY corrections compared with the $m_D=0$ case.
For example, almost half of the parameter space 
is forbidden in the case with $m_D^2/M^2 = -0.3$ by the stau mass 
constraint.
Thus we can not obtain $m_b(M_Z) \geq 2.6$ GeV for $M > m_{16}$, while 
the region with $m_b(M_Z) \geq 2.6$ GeV and $M \ll m_{16}$ becomes 
slightly wider compared with the $m_D=0$ case.
Therefore small values of $|m_D^2|$ are favorable for realistic models.
Our result is expected to agree with that in SUSY $SU(5)$ GUT models
with a large $\tan \beta$ because
the soft scalar mass spectrum in the presence of $D$-term contribution 
in SUSY $SO(10)$ GUT models has the same pattern as that in SUSY $SU(5)$
GUT.  
In fact, our result is consistent with that in Ref. \cite{softGYU2}.

To summarize, we have considered phenomenological 
implications of relations (\ref{sumrule1}) and (\ref{sumrule2}) 
within the framework of SUSY $SO(10)$ GUT.
We have investigated constraints due to successful electroweak 
symmetry breaking and the positivity of the stau mass squared.
These forbid the parameter region with small gaugino mass and 
$m_{16} < 0.4 M$.
We have also estimated the bottom mass in the allowed region.
Further, the allowed regions are more constrained by other requirements, 
e.g., the lightest superparticle should be neutral 
and this particle should not overclose our universe.
Also we have taken into account $D$-term contributions to soft scalar 
masses.
Small $D$-term contributions are 
favorable to realistic models.
It is an interesting subject to construct a realistic SUSY $SO(10)$
GUT with sum rules refering to our result.\footnote{
The gauge-Yukawa unification is studied by the use of 
explicit SUSY $SO(10)$ gauge-Yukawa models in Ref. \cite{GYUSO10}.}

\section*{Acknowledgments}  
This work was partially supported by the Academy of Finland under 
Project no. 37599 and the European Commission TMR
program ERBFMRX-CT96-0045 and CT96-0090.
The authors would like to thank J. Kubo and H.P. Nilles
for useful discussions and encouragement.

\end{document}